\documentclass[sts]{imsart}
\usepackage{amssymb}
\RequirePackage[OT1]{fontenc}
\usepackage{amsthm,amsmath,natbib}
\RequirePackage[colorlinks,citecolor=blue,urlcolor=blue]{hyperref}
\usepackage{algorithm}
\usepackage[noend]{algpseudocode}
\usepackage{graphicx}
\usepackage{tikz}
\usetikzlibrary{arrows}
\usetikzlibrary{positioning}
\newdimen\nodeDist
\makeatletter
\def\BState{\State\hskip-\ALG@thistlm}
\makeatother

\startlocaldefs

\algnewcommand\algorithmicinput{\textbf{Input:}}
\algnewcommand\INPUT{\item[\algorithmicinput]}

\algnewcommand\algorithmicoutput{\textbf{Output:}}
\algnewcommand\OUTPUT{\item[\algorithmicoutput]}

\def\Yobs{Y_{obs}}
\def\Yobsj{Y_{j,obs}}

\def\Ymis{Y_{mis}}
\def\Ymisj{Y_{j,mis}}

\def\Yimp{Y_{imp}}
\def\Yimpj{Y_{j,imp}}
\def\Yimpnotj{Y_{(-j),imp}}

\def\Ynotj{Y_{(-j)}}

\DeclareMathOperator{\E}{E}
\DeclareMathOperator{\Var}{Var}

\newcommand{\beq}{\begin{equation}}
\newcommand{\eeq}{\end{equation}}

\newcommand{\ben}{\begin{enumerate}}
\newcommand{\een}{\end{enumerate}}

\newcommand{\iid}{\overset{iid}{\sim}}

\newcommand{\Hyi}{H_i^{\mathcal{Y}}}
\newcommand{\Hxi}{H_i^{\mathcal{X}}}
\newcommand{\hy}{r}%
\newcommand{\hx}{s}%
\newcommand{\ky}{{k^{\mathcal{Y}}}}
\newcommand{\kx}{{k^{\mathcal{X}}}}
\newcommand{\kz}{{k^{\mathcal{Z}}}}
\newcommand{\byx}[1]{#1^{\mathcal{X}}}
\newcommand{\byy}[1]{#1^{\mathcal{Y}}}

\numberwithin{equation}{section}
\theoremstyle{plain}

\endlocaldefs

\begin{document}

\begin{frontmatter}
\title{Multiple Imputation: A Review of Practical and Theoretical Findings
}
\runtitle{Multiple Imputation}

\begin{aug}
\author{\fnms{Jared S.} \snm{Murray}\thanksref{t1}\ead[label=e1]{jared.murray@mccombs.utexas.edu}}%

\thankstext{t1}{The author gratefully acknowledges support from the National Science Foundation under grant numbers SES-1130706, SES-1631970 and DMS-1043903. Any opinions, findings, and conclusions or recommendations expressed
in this material are those of the author(s) and do not necessarily reflect the views of the funding agencies. }
\runauthor{J.S. Murray}

\affiliation{University of Texas at Austin}

\address{12110 Speedway B6500, Austin, Texas.\ \printead{e1}.}

\end{aug}

\begin{abstract}
Multiple imputation is a straightforward method for handling missing data in a principled fashion. This paper presents an overview of multiple imputation, including important theoretical results and their practical implications for generating and using multiple imputations.  A review of strategies for generating imputations follows, including recent developments in flexible joint modeling and sequential regression/chained equations/fully conditional specification approaches. Finally, we compare and contrast different methods for generating imputations on a range of criteria before identifying promising avenues for future research.
\end{abstract}


\end{frontmatter} 
\section{Introduction}

Multiple imputation (MI) \citep{Rubin1987} is a simple but powerful method for dealing with missing data.  MI as originally conceived proceeds in two stages: A data disseminator creates a small number of completed datasets by filling in the missing values with samples from an imputation model. Analysts compute their estimates in each completed dataset and combine them using simple rules to get pooled estimates and standard errors that incorporate the additional variability due to the missing data.  %

MI was originally developed for settings in which statistical agencies or other data disseminators provide multiply imputed databases to distinct end-users. There are a number of benefits to MI in this setting: The disseminator can support approximately valid inference for a wide range of potential analyses with a small set of imputations, and the burden of dealing with the missing data is on the imputer rather than the analyst.  All analyses conducted on the publicly available files can be based on the same set of imputations, ensuring that differences in results are not due to the handling of missing data.

With the introduction of easy-to-use software to generate imputations and combine estimates it has become increasingly common for users to create their own imputations prior to analysis.  The set of methods available to generate imputations has also grown substantially, from simple parametric models and resampling methods to iterative classification and regression tree-based algorithms and flexible Bayesian nonparametric models. There are several textbook treatments of multiple imputation (e.g. \cite{Rubin1987,LittleRubin200209,van2012flexible,CarpenterKenward201302}) but fewer recent reviews of the variety of methods available to create multiply imputed files.

This paper provides a review of MI, with a focus on methods for generating  imputations and the theoretical results and empirical evidence available to guide the selection and critique of imputation procedures.  We restrict attention to methods for imputing item missing data (imputing the subset of values that are missing for an incomplete observation) in settings with independent observations. Much of the discussion also applies to other data structures, and to problems other than item missing data where MI has proven useful (see \cite{reiter:raghu:07} for some examples of other uses for multiple imputation). 

The paper proceeds as follows: Section~\ref{sec:how} briefly reviews the mechanics of multiple imputation for a scalar estimand. Section~\ref{sec:when} reviews the conditions under which the usual MI rules give valid inference. Section~\ref{sec:implications} summarizes the practical implications of the theoretical results, particularly for choosing a method for generating imputations. Section~\ref{sec:singlegen} reviews methods for imputing a single variable subject to missingness. Section~\ref{sec:multiplegen} reviews methods for imputing several variables. Section~\ref{sec:choosing} discusses some of the considerations for choosing an imputation model. Section~\ref{sec:conclusion} concludes with discussion and directions for future work.

\section{Multiple imputation: How Does it Work?}\label{sec:how}

Let $Y_i = (Y_{i1}, Y_{i2}, \dots Y_{ip})$ denote a $p-$dimensional vector of values corresponding to the $i^{th}$ unit and $R_i = (R_{i1}, R_{i2}, \dots R_{ip})$ be a vector of indicator variables representing the response pattern, where $R_{ij} = 1$ if $Y_{ij}$ is observed and is zero otherwise.  We will use lowercase letters to distinguish fixed values from random variables, and denote the realized values in a particular dataset with a tilde (e.g., $R_i$ is a random vector, $r_i$ is a particular value that might be taken by $R_i$, and $\tilde r_i$ is the observed response pattern for unit $i$ observed in a particular dataset). 

Let $R = \{R_i : 1\leq i\leq n\}$ with $r$ and $\tilde r$ defined similarly.
The observed and missing values from a dataset of size $n$ with response pattern $R$ are denoted $\Yobs(R) = \{Y_{ij} : r_{ij}=1,\ 1\leq j\leq p,\ 1\leq i\leq n\}$ and $\Ymis(R) = \{Y_{ij} : r_{ij}=0,\ 1\leq j\leq p,\ 1\leq i\leq n\}$, respectively.
Where the explicit dependence on the response pattern is a distraction we will drop the functional notation and simply refer to $\Ymis$ and $\Yobs$.

We assume throughout that the missing data are {\em missing at random (MAR) }  \citep{Rubin1987}, that is,
\begin{align}
\Pr( R=\tilde r\mid \Yobs(\tilde r)=\tilde y_{obs}&, \Ymis(\tilde r)=y_{mis}, \phi)
\end{align}
takes the same value for all $y_{mis}$ and $\phi$, where $\phi$ parameterizes our model of the response mechanism (the distribution of $(R\mid Y)$). Under MAR we do not need to explicitly model the response process to impute the missing data. \citep[Result 2.3]{Rubin1987}. MI may be used for missing data that are not MAR provided we explicitly model the response mechanism or make other identifying assumptions (see \cite{Rubin2003-zl} for related discussion and examples of MI for non-MAR missing data).

\subsection{Multiple imputation for a scalar estimand}\label{sec:ignorable-mi}

Let $Q$ be an estimand of interest, which may be a function of complete data in a finite population or a model parameter.  Let $\hat Q(Y)$ be an estimator of $Q$ with sampling variance $U$ estimated by $\hat U(Y)$; where there is no ambiguity we refer to these as $\hat Q$ and $\hat U$. 
In order to fix ideas we focus on scalar $Q$.  Inference for vector $Q$ is similar in spirit; see \cite[Chapter 3]{Rubin1987}, also \cite[Chapter 4, Section 3]{Schafer1997} or the review in \cite[Section 2.1]{reiter:raghu:07}. 

Assume $\Ymis^{(1)}, \Ymis^{(2)}, \dots, \Ymis^{(M)}$ are $M$ imputations for $\Ymis$. %
Define $\hat Q^{(m)}= \hat Q(\Yobs, \Ymis^{(m)})$, the estimator computed using the $m^{th}$ completed dataset (with $\hat U^{(m)}$ defined similarly), and
\begin{gather}
\bar Q_M = \sum_{m=1}^M \frac{\hat Q^{(m)}}{M},\quad
\bar U_M = \sum_{m=1}^M \frac{\hat U^{(m)}}{M},\quad 
B_M = \sum_{m=1}^M \frac{(\hat Q^{(m)} - \bar Q_M)^2}{M-1}\label{eq:miB}.
\end{gather}
These statistics form the basis for inference under MI:  $\bar Q_M$ averages the estimate computed in each imputed dataset to obtain an estimate of $Q$. The variance estimator of $\bar Q_M$ has an ANOVA style decomposition:
\begin{equation}
T_M = \bar U_M + \left(1+\frac{1}{M}\right)B_M,
\end{equation}
where $\bar U_M$ is an estimate of the variance of $\hat Q$ if we had the complete data (``within-imputation'' variance), and $B_M$ estimates the excess variance due to the missing values (``between-imputation'' variance). The factor $(1+1/M)$ is a bias adjustment for small $M$, as explained in \cite[][Chapter 3.3]{Rubin1987}.

MI was originally derived under Bayesian considerations. The Bayesian derivation of MI begins with the identities
\begin{align}
P(Q\mid \Yobs) &= \int P(Q\mid \Ymis, \Yobs)P(\Ymis\mid \Yobs)\,d\Ymis\label{eq:mippred}\\
\E(Q\mid \Yobs) &= \E(\E(Q\mid \Ymis, \Yobs)\mid \Yobs)\label{eq:mibayesQ}\\
\Var(Q\mid \Yobs) &= 
\E(\Var(Q\mid \Ymis, \Yobs)\mid \Yobs)
 \nonumber\\
&\quad + \Var(\E(Q\mid \Ymis, \Yobs)\mid \Yobs)\label{eq:mibayesT}
\end{align}
When imputations are generated from $P(\Ymis\mid\Yobs)$, the MI statistics are Monte Carlo estimates of the relevant quantities: 
\begin{align}
\bar Q_M&\approx \E(\E(Q\mid \Ymis, \Yobs)\mid \Yobs) = \E(Q\mid \Yobs)\\
\bar U_M&\approx \E(\Var(Q\mid \Ymis, \Yobs)\mid \Yobs),\\
 (1+1/M)B_M&\approx \Var(\E(Q\mid \Ymis, \Yobs)\mid \Yobs)\\
 T_M &\approx \Var(\E(Q\mid \Yobs)).
\end{align}

\cite{Rubin1987} proposed constructing confidence intervals for $Q$ based on an asymptotic normal approximation to the posterior distribution \eqref{eq:mippred}: Taking $M$ to infinity, $(\bar Q_\infty-Q)\sim N(0, T_\infty)$ approximately in large samples.  In large samples with finite $M$ interval estimation for $Q$ proceeds using a reference $t-$distribution for $\bar Q_M$: $(\bar Q_M-Q)\sim t_{\nu_M}(0, T_M)$.  \cite{Rubin1987} computed an approximate value for $\nu_M$ using a moment matching argument, obtaining $\nu_M = (M-1)\left(1 + {1/r_M} \right)^2$ where $r_M = (1+1/M)B_M/{\bar U_M}$ is a measure of the relative increase in variance due to nonresponse. \cite{Barnard1999-ke} proposed an alternative degrees of freedom estimate with better behavior in moderate samples, suggesting it for general use.  See \cite{reiter:raghu:07} for a review of combining rules for more general estimands.

\section{Multiple Imputation: When Does it Work?}\label{sec:when}

In this section we give a high-level review of some of the justifications for using MI and the estimators given above. Special consideration is given to results that can inform the selection of an imputation model. 

\subsection{Bayesian (in)validity under MI}\label{sec:bayes-valid}

Since the MI estimators were derived under Bayesian arguments we might hope that MI yields valid Bayesian inference. In general it does not. Suppose the analyst has specified a Bayesian model as $P_A(Y, Q) = P_A(Y\mid Q)P_A(Q)$.  The analyst's inference is based on the posterior distribution
\begin{align}
P_A(Q\mid \Yobs) &= \int P_A(Q\mid \Ymis, \Yobs)P_A(\Ymis\mid \Yobs)\,d\Ymis. \label{eq:QAnalystpost}
\end{align}

Now suppose the imputer has generated imputations according to $\Ymis^{(m)}\sim P_I(\Ymis\mid \Yobs)$. On computing $\hat Q(\Yobs, \Ymis^{(m)})$ the analyst has a draw from the hybrid model
\begin{align}
P_H(Q\mid \Yobs) &= \int P_A(Q\mid \Ymis, \Yobs)P_I(\Ymis\mid \Yobs)\,d\Ymis\label{eq:QHybridpost}
\end{align}
If $P_A(\Ymis\mid\Yobs) = P_I(\Ymis\mid\Yobs)$, then MI delivers the analyst's posterior inference in the sense that $\hat Q^{(m)}$ is a draw from \eqref{eq:QAnalystpost}. If the posterior distribution for $Q$ is approximately normal and $M$ is not too small the MI statistics will give a reasonable approximation to the posterior.

However, in practice the imputer and the analyst will likely have different models for $(\Ymis\mid \Yobs)$. 
Even if one analyst should happen to share the same model as the imputer, the next analyst may have a different set of beliefs encoded in their model, resulting in $P_{A'}(\Ymis\mid\Yobs) \neq P_A(\Ymis\mid\Yobs)$. In this case the imputer cannot deliver valid Bayesian inference to both analysts with a single set of imputations.
Since Bayesian validity is generally unattainable (and good repeated sampling behavior is desirable in its own right), MI is usually evaluated based on its frequentist properties.  The remaining subsections explore conditions under which MI yields valid frequentist inference.

\subsection{Frequentist Validity: Conditions on complete data inference}
We will follow \cite{Rubin1996} and assume that the complete data inference is at least {\em confidence valid}, meaning that a nominal $100(1-\alpha)\%$ confidence interval has actual coverage at least  $100(1-\alpha)\%$. (The stronger condition of {\em randomization validity} requires that the nominal and actual coverage rates agree.) We also assume that the sampling distribution of $\hat Q$ is normal, so that valid confidence intervals can be obtained from $\hat Q$ and $\hat U$. In this case confidence validity requires that
\begin{align}
\E(\hat Q) &= Q\label{eq:comdata1}\\
\E(\hat U) &\geq \Var(\hat Q)\label{eq:comdata2},
\end{align}
where the expectation and variance are over repeated sampling. Randomization validity obtains when $\E(\hat U) = \Var(\hat Q)$.
We depart slightly from \cite{Rubin1996,Rubin1987} in omitting any conditioning on fixed values in a finite population.

In practice normality and \eqref{eq:comdata1}-\eqref{eq:comdata2} may only hold asymptotically, or when particular modeling assumptions are correct.  Whether this is plausible for a particular analysis will depend on the nature of $\hat Q$. For our purposes we will assume that any necessary conditions for confidence validity with completely observed data are satisfied, since our primary consideration is the impact of missingness and imputation.  Of course, if the complete data inference is not valid it would be unreasonable to expect MI or any other missing data procedure to remedy the issue.

\subsection{Proper imputation for valid inference}

Chapter 4, Section 4.2 in \cite{Rubin1987} outlines conditions under which MI inferences are randomization or confidence valid when $M=\infty$. Imputations satisfying these conditions for a particular estimand $Q$ and posited response mechanism are known as {\em proper} imputations. Proper imputation coupled with valid complete data inference yields valid MI inference \citep[Result 4.1]{Rubin1987}. It is important to remember that imputations are only proper with respect to a particular estimand $Q$ and a posited response mechanism. 

We focus on three essential conditions necessary for an imputation procedure to be proper for an estimand $Q$. (The other conditions are somewhat technical and generally not the source of improper imputations and invalid inference in practice.)

\subsubsection{Three essential conditions for proper imputation.}\label{sec:essential}

\cite{Rubin1996} distilled the formal definition of proper imputation given in \cite[Section 4.2]{Rubin1987} into three conditions that generally ensure imputations are proper. They concern the behavior of the MI statistics under repeated realizations of the response mechanism, holding the sample values $Y$ fixed (that is, under repeated sampling from $P(R\mid Y)$). The first two conditions require that $\bar Q_{\infty}$ and $\bar U_{\infty}$ be approximately unbiased for $\hat Q$ and $\hat U$:
\begin{align}
\E(\bar Q_\infty \mid Y) &\approx \hat Q(Y)\label{eq:prop1} \\
\E(\bar U_\infty \mid Y) &\approx \hat U(Y)\label{eq:prop2},
\end{align}
where the expectations are with respect to $P(R\mid Y)$. 

Naturally \eqref{eq:prop1}-\eqref{eq:prop2} will hold if $P(\Ymis\mid\Yobs)$ is correctly specified by the imputer. However, imputations made under misspecified models can still satisfy \eqref{eq:prop1}-\eqref{eq:prop2} so long as they broadly capture the features of the predictive distribution that are relevant for computing $Q$ and $U$ and the proportion of missing data is not extreme. To see this more clearly we can write %
\begin{align}
\E(\bar Q_\infty \mid Y) &= \sum_{m=1}^\infty \E\left( \hat Q(Y_{obs}(R), Y^{(m)}_{mis}(R))\mid Y\right).
\end{align}
With no missing data the expectations inside the sum are all $\hat Q(Y)$. With modest amounts of missing data, the imputed values need to be sufficiently poor to overwhelm the influence of the observed data in computing $Q$. (What constitutes ``sufficiently poor'' naturally depends on $Q$.) Similar logic applies to $\bar U_\infty$.

The third condition for proper imputation is more subtle: It requires that the between-imputation variability $B_\infty$ be approximately unbiased for the variance of $\bar Q_\infty$:
\begin{equation}
\E(B_\infty\mid Y)\approx \Var(\bar Q_\infty\mid Y).\label{eq:properB}
\end{equation}
Satisfying this condition generally requires that we account for uncertainty {\em in the imputation model itself} (or equivalently uncertainty in the parameters indexing a model class), since the observed data used to estimate the model, $\Yobs(R)$, varies over samples from the response mechanism. (Recall that the variance in \eqref{eq:properB} is with respect to $P(R\mid Y)$.) 

Many seemingly reasonable stochastic imputation procedures fail to be proper because they do not satisfy \eqref{eq:properB}; these include imputing from a model by plugging in the MLE or drawing imputations from the empirical distribution of observed cases \citep[][Ch. 4]{Rubin1987}.  Accounting for uncertainty in the imputation model can be achieved (or approximated) in a variety of ways, such as sampling the parameters indexing a particular model class from their posterior under a Bayesian model or through small adjustments to the bootstrap (as described in Section~\ref{sec:hotdeck}). See Section~\ref{sec:uncertainimp} for further discussion.

\subsection{Congeniality and confidence validity}\label{sec:congenial}

It is well-known that the MI estimate $T_\infty$ can be inconsistent for certain choices of $Q$ \citep{Wang1998-wp, Robins2000-wp,Kim2002-cy, Nielsen2003-er,  Kim2006-ea}. The bias is typically positive and tends to have limited influence on coverage rates for common estimands when the amount of missingness is not extreme \citep{Rubin2003-zl}.  \cite{Rubin1996} reviewed early examples of inconsistency and gave sufficient conditions for MI inference to be confidence proper (i.e., for $T_\infty$ to conservatively estimate $Var(\bar Q_\infty)$); they are similar to the conditions in Section~\ref{sec:essential}, averaged over repeated sampling of $Y$ in addition to the response mechanism. %

\cite{Meng1994} introduced the concept of {\em congeniality} for understanding the inconsistency of the MI variance estimate.  Roughly, an analysis procedure is { congenial} to an imputation model $P_I(\Ymis\mid\Yobs)$ if we can take the complete data analysis and embed it into a Bayesian model $P_A(Y\mid Q)P_A(Q)$ such that
\begin{enumerate}
\item Its posterior $P_A(Q\mid Y)$ recapitulates the desired analysis in the sense that
\begin{gather}
\E_A(Q\mid Y) = \hat Q(Y),\quad \Var_A(Q\mid Y) = \hat U(Y).
\end{gather}
\item It matches the imputation model, i.e.,
\begin{equation}
P_A(\Ymis\mid \Yobs) = P_I(\Ymis\mid \Yobs).
\end{equation}
\end{enumerate}
Under congeniality, MI delivers samples from $P_A(Q\mid \Yobs)$ (Section~\ref{sec:bayes-valid}), which we have constructed to yield confidence valid inference. Unless the analyst is the imputer, congeniality is less a condition we should try to satisfy than one we should try to fail gracefully -- uncongeniality is generally ``the rule not the exception'' \citep{Xie2017congenial}, for the same reasons discussed in Section~\ref{sec:bayes-valid}.

\cite{Xie2017congenial} revisited the behavior of MI inferences under uncongeniality and provided a host of new results.  At a high level their findings affirm and generalize common rules of thumb originating with \cite{Meng1994}: Even if the ``true'' model is nested within the imputer's and the analyst's models (e.g., if the imputation model includes both relevant and irrelevant covariates in an otherwise correctly specified regression model for the missing data), standard MI inference may be invalid. However, if the analyst's procedure is {\em self-efficient} (meaning essentially that their estimator cannot be improved by ignoring relevant data \citep{Meng1994,Meng2003-qm}), then:
\begin{enumerate}
\item When the imputer's model is more saturated than the analyst's, the usual MI inference is confidence valid and generally robust.
\item When the imputer's model is less saturated than the analyst's, confidence validity is not guaranteed.
\end{enumerate} 
It is generally safer to conduct an uncongenial analyses under (1) than under (2), since conservative inferences will obtain. \cite{Xie2017congenial} also provide remarkably simple and broadly applicable (if somewhat exacting) alternative variance estimates that are valid under uncongeniality: Use $T^*_M=2T_M$ for a vector $Q$, or sum and square the standard errors for a univariate $Q$: $T^*_M = (\sqrt{U_M} + \sqrt{B_M})^2 + (1/M)B_M$. 

Like most strong theoretical results, \cite{Xie2017congenial}'s results depend on a number of assumptions. One of these assumptions is that the true model (``God's model'') is nested within the imputation model class. In his discussion of the paper, \cite{Reiter2017congenial} notes that ``[I]n my experience, very low coverage rates in MI confidence intervals arise more often from the imputation procedure generating bias in [$\bar Q_\infty$] than from bias in the MI variance estimator,'' often due to rote application of default imputation procedures. This has been in part a shared experience (\cite{Murray2016-cy}), motivating the focus of this review on the specification of imputation models.

\section{Practical Implications of theoretical results for Imputation Modeling}\label{sec:implications}

The theoretical results summarized above suggest a number of practical considerations for generating imputations.  These are reviewed below; for more detailed discussion and examples, see e.g. \cite{Rubin1987,Little1988,Rubin1996,van2012flexible}. Throughout this section and the rest of the paper we will continue to refer to procedures that generate imputations as ``imputation models'', regardless of whether they are completely specified probability models.

\subsection{Imputations should reflect uncertainty about missing values and about the imputation model.}\label{sec:uncertainimp}

The goal in multiple imputation is to account for uncertainty due to the missing values in subsequent inference. This is a different objective than estimating or predicting the missing values, which could generally be achieved via simpler means. The situation in MI is similar to the more familiar task of constructing valid predictive intervals with a regression model, where we need to account for uncertainty in the unobserved response as well as uncertainty in the regression fit. 

Suppose we have a single variable subject to missingness, to be imputed using a regression model. If we were only concerned with reconstructing the missing values, we would just impute the fitted values. This would clearly lead to invalid MI inferences. Instead, MI propagates the intrinsic uncertainty about the missing values via some stochastic mechanism, for example, by adding a randomly generated residual to the regression prediction. However, to achieve at least approximately proper imputations we also need to account for uncertainty about the imputation model itself -- that is, uncertainty in the fitted values of the regression model. 
Methods that do not appropriately reflect both sources of uncertainty tend to violate \eqref{eq:properB} and underestimate the between-imputation variance, yielding standard errors that are too small and anti-conservative inferences \citep{Rubin1987,Rubin1996}. 

Bayesian imputation procedures provide a natural mechanism to account for model uncertainty. Imputations are generated from
\begin{equation}
P(\Ymis\mid\Yobs) = \int P(\Ymis\mid \theta, \Yobs)P(\theta\mid\Yobs)d\theta.
\end{equation}
where $\theta$ is a parameter indexing a model for $Y$ (or a model for $\Ymis$ given $\Yobs$). To see how model uncertainty propagates, observe that imputations can be sampled compositionally: For $1\leq m\leq M$, first draw a value $\theta^{(m)}\sim P(\theta\mid\Yobs)$ and then sample $\Ymis^{(m)}\sim P(\Ymis\mid \theta^{(m)}, \Yobs)$. Model uncertainty is represented by $P(\theta\mid\Yobs)$, and the intrinsic uncertainty about the missing values is represented by $P(\Ymis\mid \theta, \Yobs)$.
Approximations to full Bayesian inference have also proven useful: \cite{Rubin1986-ge}'s approximate Bayesian bootstrap for proper hot deck imputation is one early example (Section~\ref{sec:hotdeck}). Chapter 10 of \cite{LittleRubin200209} reviews several others. 

Of course, Bayesian modeling is not magic -- if $\theta$ indexes a class of misspecified models then we should expect our imputations and inferences to suffer, at least for estimands that are sensitive to this misspecification.  For example, when $\Ymis$ contains variables with significant skew a multivariate normal imputation model would likely yield approximately valid inference for marginal means but invalid inference for some marginal quantiles, since \eqref{eq:prop1} can be violated when $Q$ is an extreme quantile. 

From a coverage perspective, model misspecification becomes increasingly consequential in large samples where the complete data standard errors are small and $P(\theta\mid\Yobs)$ will tend to concentrate on the parameters of the ``best'' misspecified model.  Even small biases due to misspecification in the imputation model can become large relative to the pooled standard errors. Enlarging the imputation model class $P(Y \mid \theta)$ via non- and semiparametric Bayesian modeling can guard against misspecification and also mitigate the artificial certainty implied by fixing a regular parametric model and only considering uncertainty in its parameters.  Section~\ref{sec:npbayesimp} explores recent promising developments in this area.

\subsection{Imputation models should generally include as many variables as possible.}

There are multiple reasons for entertaining the largest possible imputation model: The missing at random assumption tends to be more tenable as more completely-observed variables are added to the imputation model.  In addition, if variables predictive of the missing values are left out of the imputation model but used to compute $Q$ or $U$, then the imputations will be improper -- the imputed values will be incorrectly independent of the omitted variables, leading to bias over repeated imputations (violations of \eqref{eq:prop1} or \eqref{eq:prop2}) \citep{Rubin1996}.  In this case the analysis and imputation models are uncongenial in the ``wrong'' way -- the imputer's model is less-saturated than the analysis model.  In sum, the cost of excluding a relevant variable (invalid inference) is often greater than the cost of including an irrelevant variable (roughly, additional variance).  This is particularly relevant when the analyst and imputer are not the same, and the imputations must support many unspecified analyses.  Even when the imputer and the analyst are the same it would be useful to generate one set of imputations that can support the usual process of iterative model building and refinement, rather than generating a new set of imputations for each analysis model that is considered.  See \cite{Collins2001-bv} and \cite{Schafer2003-hh} for further discussion of the tradeoffs involved.

These points are particularly relevant for design variables in complex surveys.   Design-based estimators will typically use stratum and cluster information to compute $U$.  \cite{Reiter2006-za} show empirically that failing to account for an informative sampling design can lead to invalid inference. They suggest including indicator variables for strata and cluster membership in the imputation model, or including stratum fixed effects and cluster random effects in imputation models. It may be useful to include estimated response propensities or final adjusted survey weights (sampling weights with e.g. calibration and post-stratification adjustments) as well, especially if complete design information is not available to the imputer \citep{Rubin1996}.  %

\subsection{Imputation models should be as flexible as possible.}
Finally, imputation models should try to ``track the data'' \citep{Rubin1996} by modeling relevant features of the joint distribution of the missing values. Loosely, a feature of the joint distribution is relevant if it is a possible target of inference itself, or more generally if it yields a more accurate predictive distribution for the missing data.  Interactions, nonlinearities, and non-standard distributional forms are all potentially relevant features.

As \cite{Meng1994} succinctly put it, ``Sensible imputation models should not only use all available information to increase predictive power, but should also be as general and objective as practical in order to accommodate a potentially large number of different data analyses.''  We would add that where possible, imputation models should have some capacity to {\em adapt} to unanticipated features of the data (such as interactions, nonlinearities, and complex distributions), especially when the imputer has limited time and resources to spend on iteratively improving the imputation model. %

\section{Generating imputations for a single variable}\label{sec:singlegen}

We begin by cataloging some of the more common approaches to generating imputations for a single variable subject to missingness, conditional on other fully observed variables. In the next section we consider how these can be extended to generate imputations for several variables.

\subsection{Regression Modeling}

Imputation by sampling from univariate regression models is conceptually straightforward. Generalized linear models and extensions to deal with complications such as zero-inflation and truncation are popular options; these are not reviewed in depth here but see e.g. \cite{VanBuuren1999}, \cite{Raghunathan2001}, \cite{su2011multiple}, or \cite{van2012flexible} (Chapter 3). These methods are quite common in practice, but since most readers will be familiar and they are well-reviewed elsewhere we will not enumerate them here.

To generate proper imputations some method should be used to account for parameter uncertainty -- simple strategies like sampling from the regression model with parameters fixed at the observed data MLE are generally improper. 
Posterior sampling under a non- or weakly informative prior tends to be proper when the model fits well. Prior distributions can also ease problems like separation in logistic regression and apply helpful regularization in conditional models with many variables in the conditioning set \citep{su2011multiple}.

\subsection{Hot Deck/Nearest Neighbor Methods.}\label{sec:hotdeck}

The hot deck and other nearest-neighbor methods \citep{Chen2000-gy,Andridge2010-hu} begin by defining a distance metric between cases in terms of the observed covariates. Imputations for a missing value are borrowed from a nearby completely observed case (the ``donor''). These methods tend to be simpler to implement than fully specified regression models and often make fewer assumptions. However, these methods are far from assumption free -- the choice of distance metric, the definition of the donor pool, and how to sample from the donor pool all influence the quality of imputations.

The hot deck \citep{Andridge2010-hu} defines distance via cross-classifications of fully observed variables which determine adjustment cells. Missing values are imputed by sampling with replacement from the pool of donors within the same cell. This strategy ensures that all imputations are plausible values, which is an appealing feature relative to regression imputation. Complications arise when there are many fully observed variables to incorporate into the cross-classification or when the sample size is low, leading to many small or empty adjustment cells. 

MI with the hot deck is also known to be improper for simple estimands like a population mean \citep{Rubin1986-ge}. The hot deck effectively assumes that the distribution of missing values within an adjustment cell is {\em exactly} the empirical distribution of the observed values within that cell, which leads to $B$ having downward bias (due to ignoring uncertainty in the implicit imputation model). \cite{Rubin1986-ge} propose a simple modification that makes the hot deck proper, based on an approximation to the Bayesian bootstrap \citep{Rubin1981-os}. Instead of sampling the $n_m$ missing values from the empirical distribution of the $n_o$ observed values within an adjustment cell, the approximate Bayesian bootstrap (ABB) first samples a set of $n_o$ values with replacement from the observed data and then samples $n_m$ imputed values with replacement from this set. This simple adjustment yields proper imputations for the population mean of the adjustment cell \citep{Rubin1986-ge}. (See also \cite{Kim2002-cy} for a more accurate variance estimate in small samples.)

Predictive mean matching (PMM) \citep{Little1988} instead measures the distance between cases by the distance between their predicted means for the variable subject to missingness (traditionally estimated using a linear regression, although in principle any method could be used to make the prediction).  PMM generalizes the hot deck, which is a special case of PMM using saturated models with categorical predictors. By avoiding the discretization and making some assumptions about the relationships between the predictors and the response (such as linearity) PMM can handle more variables than the hot deck, but may be sensitive to the predictive model specification.

To define the donor pool \cite{Heitjan1991-jn} proposed sampling from a window of $k$ nearby potential donors in PMM in the hope of making the method approximately proper.  The donor's value may be imputed, or its residual can be added to the predicted mean of the missing value to generate an imputation.    \cite{Schenker1996-bp} found these two approaches to perform similarly in simulations; the former will always impute a previously realized value, which may be desirable.  See \cite{Vink2014-yr} for an approach to semi-continuous variables.  \cite{Morris2014-vx}  compared newer developments and current implementations of these techniques, cautioning in particular against the imputation of a single nearest neighbor (which appears to be common in software implementations of PMM) as it is improper.

\begin{figure}[ht]
\begin{center}
\begin{tikzpicture}[
  scale=0.9,
    node/.style={%
      draw,
      rectangle,
    },
    node2/.style={%
      draw,
      circle,
    },
  ]

    \node [node] (A) {$y_1<0.9$};
    \path (A) ++(-135:\nodeDist) node [node2] (B) {$A_1$};
    \path (A) ++(-45:\nodeDist) node [node] (C) {$y_2<0.4$};
    \path (C) ++(-135:\nodeDist) node [node2] (D) {$A_2$};
    \path (C) ++(-45:\nodeDist) node [node2] (E) {$A_3$};

    \draw (A) -- (B) node [left,pos=0.25] {no}(A);
    \draw (A) -- (C) node [right,pos=0.25] {yes}(A);
    \draw (C) -- (D) node [left,pos=0.25] {no}(A);
    \draw (C) -- (E) node [right,pos=0.25] {yes}(A);
\end{tikzpicture}
\hspace{0.05\linewidth}
\begin{tikzpicture}[scale=3.3]
\draw [thick, -] (0,1) -- (0,0) -- (1,0) -- (1,1)--(0,1);
\draw [thin, -] (0.8, 1) -- (0.8, 0);
\draw [thin, -] (0.0, 0.4) -- (0.8, 0.4);
\node at (-0.1,0.4) {0.4};
\node at (0.8,-0.1) {0.9};
\node at (0.5,-0.2) {$y_1$};
\node at (-0.3,0.5) {$y_2$};
\node at (0.9,0.5) {$A_1$};
\node at (0.4,0.7) {$A_2$};
\node at (0.4,0.2) {$A_3$};
\end{tikzpicture}
\end{center}
\caption{(Left) An example CART tree, with internal nodes labeled by their splitting rules and terminal nodes given labels $A_h$. (Right) The corresponding partition of $(Y_1, Y_2)$.}
\label{fig:treestep}
\end{figure}
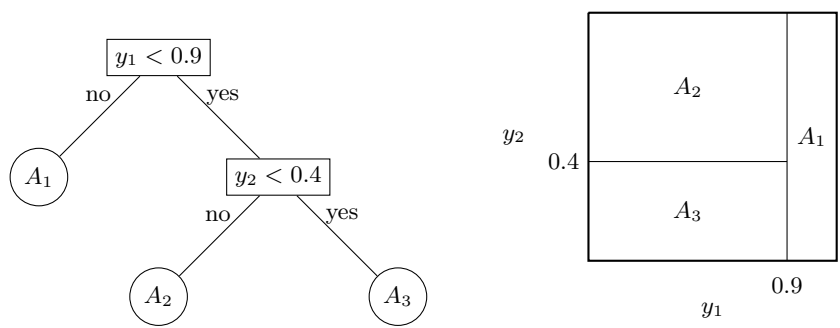

PMM and the hot deck can be made more adaptive using recursive partitioning. 
\cite{reiter2005using} and \cite{Burgette2010-dq} proposed imputation via classification and regression trees (CART, \cite{Breiman1984-jg}). A tree is grown using fully observed data to predict the variable subject to missingness. Then each incomplete case is assigned to its corresponding leaf, and an imputation is sampled from donors within in the same leaf.  The imputer can control the size of the donor pool by growing the tree down to a specified minimum leaf size.  This is a special case of PMM using CART to generate predictions; we could also think of it as an adaptive hot deck that leverages the most predictive variables and balances the size of the adjustment cells. Figure~\ref{fig:treestep} shows an example tree grown on two variables $(Y_1, Y_2)$ to impute a third ($Y_3$), along with the corresponding partition which forms the adjustment cells.

\cite{reiter2005using} and \cite{Burgette2010-dq} drew ABB samples from within the leaves in an effort to generate proper imputations.  \cite{van2012flexible} (Algorithm 3.6) suggested also accounting for uncertainty in the tree itself by growing it on a different bootstrap sample for each imputed dataset. \cite{Doove2014-na} proposed imputation by growing a random forest (an ensemble of trees) \citep{breiman2001random} of size $k$ by bootstrapping the complete cases and (optionally) sub-sampling the variables, as in traditional applications of random forests. An imputed value is generated by sampling from the $k$ trees and then following the procedure to generate a CART imputation. \cite{Shah2014-ul} proposed fitting a random forest, estimating its predictive error variance, and generating imputations as the random forest prediction plus a normally distributed residual.  

Limited results exist comparing these different recursive partitioning methods, and there is similarly limited guidance as to how they should be tuned. But they can be fast and effective imputation engines, particularly for large sets of categorical variables that take a relatively limited set of levels (see e.g. \cite{Akande2017-gq}).

\section{Generating Imputations for Multiple Variables}\label{sec:multiplegen}

There are two basic strategies for imputing multivariate missing data: Jointly modeling the variables subject to missingness, or specifying a collection of univariate conditional imputation models that condition on all the other variables (this approach goes under various names including sequential regression multivariate imputation \citep{Raghunathan2001} and multiple imputation by chained equations \citep{VanBuuren1999}, but we will use ``fully conditional specification'' (FCS) as in \cite{VanBuuren2006}). Joint models can be further classified into ``simultaneous'' approaches that define a multivariate distribution $f(Y)$ directly or ``sequential'' approaches that build up a multivariate distribution using a ladder of conditional distributions, where the model for each variable conditions only on those earlier in the sequence. Appendix~\ref{sec:software} has pointers to software implementations of many methods described in this section.

To describe the different approaches we need some new notation: Let $\Yobsj$ and $\Ymisj$ denote the set of observed and missing values for the $j^{th}$ variable. Let $\Yimp$ denote an imputed dataset, and $\Yimpj$ denote a set of imputations for $\Ymisj$. We will use the subscript $(-j)$ to denote the same quantities for all but the $j^{th}$ variable.

\subsection{Joint specification: Simultaneous approaches}\label{sec:jointsim}

Early simultaneous joint modeling approaches were based on the multivariate normal (MVN) or $t$ distribution; these are reviewed in \cite{Schafer1997} and \cite{LittleRubin200209}. For high dimensional continuous observations low-rank structure can be imposed on the covariance matrix \citep{Audigier2016-va}. Various authors have proposed imputing categorical data under a misspecified MVN model, either leaving the continuous imputations for discrete variables as-is or rounding them based on some thresholds \citep{Horton2003-ak,Bernaards2007-ae}. This is naturally more complicated when the discrete variables are not ordinal, particularly if they take many levels. Additionally, end users may not trust imputations from a data disseminator if the imputed data appear invalid. Therefore it is often preferable to use models that are appropriate for the types of variables at hand. 

For small numbers of strictly discrete variables a simple multinomial model may be feasible. However, with a large number of discrete variables it is impossible to fit saturated multinomial models and further restrictions are necessary. Options include log-linear models \citep{Schafer1997}, latent class models \citep{Vermunt2008,Gebregziabher2010,Vidotto2015-zq}, or multiple correspondence analysis \citep{Audigier2017-ye} (which is closely related to a certain class of multivariate logit models \citep{Fithian2017-nt}). %

Joint models for mixed continuous and categorical data are also available. For the remainder of Section~\ref{sec:jointsim}, suppose we have collected the continuous variables into a vector $Y$ and the discrete variables into another vector $X$.
The
general location model (GLOM) %
\citep{Olkin1961,Little1985,Schafer1997} %
assumes that $(Y\mid X=x)\sim
N(\mu_x, \Sigma_x)$ and $X\sim \pi$.
 (\cite{Liu1998} generalized the $(Y\mid X)$ model to the larger class of elliptically symmetric distributions.)
The number of parameters in this saturated model grows rapidly with the sample space of $X$, so imputers typically impose further constraints. Examples include common covariance structure ($\Sigma_x\equiv \Sigma$ for all $x$), removing higher-order effects from the conditional means by specifying
$\mu_x = D(x)B$ for a matrix of regression coefficients $B$ and design
vector $D(x)$, and imposing log-linear constraints on $\pi$ to rule our higher-order interactions in the marginal model for $X$.

\subsubsection{Mixtures and Nonparametric Bayesian Models. }\label{sec:npbayesimp} Even without additional parameter constraints, most parametric joint models make restrictive assumptions. Mixture models provide a simple and expressive way to enrich a parametric model class. For example, latent class models for categorical data are mixtures of independence models (log-linear models with only main effects) which have proven useful in multiple imputation \citep[e.g.,][]{Vermunt2008, Gebregziabher2010}. Mixtures of multivariate normal distributions can model complex features of joint continuous distributions \citep{Bohning2007,Elliott2007}. 

Several Bayesian nonparametric models have recently been proposed for multiple imputation. Most of these are based on infinite mixture models or their truncated approximations (but see \cite{Paddock2002-os} for an early exception based on Polya trees, and also the sequential regression approach in \cite{Xu2016-xy}). Relative to parametric Bayesian approaches these models are appealing for their ability to grow in complexity with increasing sample size. Under some circumstances this can allow the model to capture unanticipated structure like interactions and nonlinear relationships or nonstandard distributions, reflecting these in the imputed values.

Recall that we have separated the data into vectors of categorical variables $X$ and continuous variables $Y$. For imputing multivariate categorical data, \cite{Si2013} adopt a truncated version of the Dirichlet process mixture of product multinomials (DP-MPMN) proposed
by \cite{Dunson2009}. This is a latent class model with a large number of classes (say $\kx$) and a particular prior over the class distribution. 

Suppose the $j^{th}$ categorical variable takes (possibly unordered) values indexed by ${1,2,\dots,d_j}$ and
let $\Hxi \in \{1, \dots, \kx\}$ be a latent mixture component index for observation $i$. Let $\Pr(X_{ij} = x_{ij} \mid \Hxi = \hx) = \psi_{\hx x_{ij}}^{(j)}$. The DP-MPMN model assumes that
\begin{gather}
\Pr(\Hxi=\hx) = \byx{\phi_\hx}\label{eq:dx0}\\
 \Pr(X_i = x_i \mid \Hxi = \hx, \Psi) = \prod_{j=1}^{p}\psi_{\hx x_{ij}}^{(j)}\label{eq:dx},
\end{gather}
so that the elements of $X$ are conditionally independent given the latent class membership.
The prior on $\phi$ is a truncated version of the stick-breaking construction for the Dirichlet process (DP) \citep{Sethuraman1994}, introduced in \cite{Ishwaran2001} to simplify Gibbs sampling in DP mixture models:
\begin{gather}
\byx{\phi_\hx} = {\xi_\hx}\prod_{l<\hx}(1-{\xi_l}),\quad 
\{{\xi_\hx}\}_{\hx=1}^{\kx} \iid Beta(1, {\alpha}),\quad 
{\xi_{\kx}}\equiv 1.\label{eq:Hip}
\end{gather}

The model is completed with prior distributions on $\Psi$ and $\alpha$ (see \cite{Si2013} for a complete specification). \cite{Manrique-Vallier2014-ct,Manrique-VallierSMTrunc} extended this model to assign zero probability to impossible values of $X$, such as cells that are logically impossible (pregnant men or children collecting retirement benefits) or necessarily empty due to skip patterns. \cite{Manrique-Vallier2016-zr} introduced a variant of this model for edit-imputation that simultaneously  accounts for missing values and observed values that are logically impossible but present due to measurement error. \cite{Hu2017-mc} extended this model to nested data structures (i.e., hierarchical structures like individuals nested within households) in the presence of structural zeros.

For imputing continuous data \cite{Kim2014-kz} suggested a truncated DP mixture of multivariate normal distributions. Let $\Hyi$ be the mixture component index for record $i$. This model assumes that 
\begin{gather}
\Pr(\Hyi=\hy) = \byy{\phi_{\hy}}\label{eq:ymodel0}\\
\left(Y_i\mid \Hyi=\hy, -%
\right) \sim N(\mu_\hy, \Sigma_\hy),\label{eq:ymodel}
\end{gather}
with a prior on $\byy{\phi_{\hy}}$ defined via a stick-breaking process similar to \eqref{eq:Hip}. \cite{Kim2014-kz} modified the model in \eqref{eq:ymodel} to constrain the support of $Y$ to a set $\mathcal{A}$ with bounds determined by a set of linear inequalities, so that $\Pr(Y\not\in \mathcal{A})=0$ under the prior. \cite{Kim2015-eq} extended this approach to simultaneous edit-imputation, generating imputed values for observations outside of $\mathcal{A}$ via a measurement error model.

\cite{Murray2016-cy} built a hierarchical mixture model for mixed continuous and categorical observations by combining the models in \eqref{eq:dx0}-\eqref{eq:dx} and \eqref{eq:ymodel0}-\eqref{eq:ymodel}, with two important adjustments. First, \eqref{eq:ymodel} is modified to include a regression on $X$ with component-specific coefficients:
\begin{equation}
(Y_i\mid X_i = x_i, \Hyi=\hy,-) \sim N(D(x_i)B_\hy, \Sigma_\hy).\label{eq:ymodelwithx}\\
\end{equation}
By default the design matrix $D(x_i)$ encodes main effects. Allowing the component means to depend on $X$ greatly reduces the number of mixture components necessary to capture $X-Y$ relationships. Second, the mixture component indices in each model are given a hierarchical prior introduced by \cite{Banerjee2013}:
\begin{gather}
\Pr(\Hxi=\hx, \Hyi=\hy \mid Z_i=z) 
 = \byx{\phi_{z\hx}}\byy{\phi_{z\hy}}\label{eq:hmodel}\\
 \Pr(Z_i=z) = \lambda_z,\label{eq:zmodel}
\end{gather}
Here $\lambda_z$ is assigned a stick-breaking prior, Each pair $\byx{\phi_z} = \left(\byx{\phi_{z1}},\dots,\byx{\phi_{z\kx}}\right)'$ and
$\byy{\phi_z} = \left(\byy{\phi_{z1}}, \dots,\byy{\phi_{z\ky}}\right)'$ are probability vectors also assigned independent truncated stick breaking priors. 
This is a ``mixture of mixtures'' model; marginalizing over the latent variables the joint density is
\begin{equation}
f(X_i, Y_i) = \sum_{z=1}^{\kz} \lambda_z\left(\sum_{\hy=1}^\ky \byy{\phi_{z\hy}}N(Y_i; D(X_i)B_{\hy}, \Sigma_{\hy}) \right)\left(\sum_{\hx=1}^\kx \byx{\phi_{z\hx}}\prod_{j=1}^{p}\psi_{\hx X_{ij}}^{(j)}\right)\label{eq:xy}.
\end{equation}
Each mixture component is itself composed of two mixture models, one for $(Y\mid X)$ and one for $X$. These lower-level mixtures share some parameters ($B, \Sigma,$ and $\Psi$), enforcing a degree of parsimony. 

\cite{DeYoreo2016} used a similar hierarchical mixture model constructed based on different considerations, splitting the variables into sets based on their type (ordinal or nominal) and high or low rates of missing values. An expressive model class is specified for the variables with high rates of missing values, and a simpler model class is utilized for variables with low rates of missingness. Ordinal variables are explicitly modeled as such by thresholding mixtures similar to \eqref{eq:ymodelwithx}.

Further extensions, combinations, and enhancements of these models are possible. Despite their complexity, all of these models have been shown to perform well for MI with real, complicated data and little or no tuning. 

\subsection{Fully Conditional Specification}

FCS avoids explicit joint probability models by specifying a collection of univariate conditional imputation models instead \citep{VanBuuren1999,Raghunathan2001}. Each univariate model typically conditions on all the remaining variables.  In FCS the missing values are imputed by iteratively sampling from these conditional models:
\begin{enumerate} 
\item Begin by filling in $\Ymis$ with plausible values to generate an initial completed dataset, stored in $\Yimp$ %
\item For $1\leq j\leq p$, use a univariate imputation method to sample new imputed values for $\Ymisj$ from a distribution $P(\Ymisj \mid \Yobsj, \Yimpnotj)$, and store them in $\Yimpj$.
\item Iterate the previous step until apparent convergence and return the final value of $\Yimp$
\end{enumerate}
This process is repeated $M$ times, saving the returned value as one of the $M$ imputations. Any of the univariate imputation methods in the previous section could be used. This lends FCS some flexibility relative to the joint-simultaneous approaches described above. 

But this flexibility comes at a cost: Even if each $g_j$ is a completely specified probability model, taken together they often do not correspond to a proper joint distribution for $Y$ \citep{Arnold1989-wz,Arnold2001-pf}.  A set of full conditional distributions that do not correspond to any joint distribution is said to be {\em incompatible}.  Simple adjustments like adding polynomial terms or interactions to univariate regression models can induce incompatibility \citep{Liu2014-md}.  

While the algorithm above looks like a standard Gibbs sampler, if the conditional models are incompatible the behavior of the FCS imputation algorithm is unclear: The imputations from the FCS algorithm given above may converge to a unique limiting distribution, or fail to converge to any unique limiting distribution, or converge to different distributions depending on the initial values and/or order of the updates.
\cite{Li2012} give examples of incompatible FCS models with fixed parameters whose imputations either diverge or converge to different stationary distributions depending on the order of their updates. This phenomenon seems to be rare in real data, and \cite{Zhu2015-fn} note that estimating rather than fixing parameters ameliorates at least some of the problems in \cite{Li2012}'s examples. 

There are some limited convergence results available when the fully conditional specification comprises univariate Bayesian regression models. \cite{Liu2014-md} study an iterative FCS imputation procedure that uses a set of Bayesian regression models $g_j(Y_{ij} \mid , \Ynotj, \theta_j)$ with prior distributions $\pi_j(\theta_j)$.  With a slight abuse of notation, define
\begin{align}
g_j(\Yobsj\mid \Yimpnotj, \theta_j) &= \prod_{i=1}^n g_j(Y_{ij}\mid \Yimpj, \theta_j)^{R_{ij}}\label{eq:gj1}\\
g_j(\Yimpj\mid \Yobsj, \Yimpnotj, \theta_j) &= \prod_{i=1}^n g_j(Y_{ij}\mid \Yimpnotj, \theta_j)^{1-R_{ij}} .\label{eq:gj2}
\end{align}
Algorithm~\ref{alg:fcs} gives one iteration of an iterative FCS sampler under these models.

\begin{algorithm}
\caption{Iterative FCS Sampler from \cite{Liu2014-md}}\label{alg:fcs}
\begin{algorithmic}
\item[] For $1\leq j\leq p$,
\begin{enumerate}
\item Sample $\theta_j\sim \pi_j(\theta_j\mid \Yobsj, \Yimpnotj)\propto g_j(\Yobsj\mid \Yimpnotj, \theta_j)\pi_j(\theta_j)$
\item Sample $\Yimpj\sim g_j(\Yimpj\mid \Yobsj, \Yimpnotj, \theta_j)$
\end{enumerate}
\end{algorithmic}
\end{algorithm}

We can compare this approach to a proper MCMC algorithm under a joint model.  Specifically we consider a collapsed Gibbs sampler \citep{Liu1994-ri} that targets $P(\Ymis\mid\Yobs) = \int P(\Ymis, \theta\mid \Yobs)d\theta$ directly, by jointly sampling $(\Ymisj, \theta\mid \Yobsj, \Yimpnotj)$ at each step. It is impractical to use directly, but it is helpful to make comparisons with Algorithm~\ref{alg:fcs}.

Let the joint model be given by $f(Y_i\mid \theta)$, with full conditionals $f_j(Y_{ij} \mid \Ynotj, \theta)$ and joint prior distribution $\pi(\theta)$ (where $\theta=(\theta_1,\theta_2,\dots,\theta_p)$). Define $f_j(\Yobsj\mid \Yimpnotj, \theta)$ and $f_j(\Yimpj\mid \Yobsj, \Yimpnotj, \theta)$ as in equations \eqref{eq:gj1}-\eqref{eq:gj2}. Algorithm~\ref{alg:collapsedgibbs} gives one iteration of the collapsed Gibbs sampler.

\begin{algorithm}
\caption{Collapsed Gibbs Sampler for a Joint Model}\label{alg:collapsedgibbs}
\begin{algorithmic}
\item[] For $1\leq j\leq p$,
\begin{enumerate}
\item Sample $\theta\sim \pi(\theta\mid \Yobsj, \Yimpnotj)\propto f_j(\Yobsj\mid \Yimpnotj, \theta)\pi(\theta)$
\item Sample $\Yimpj\sim f(\Ymisj \mid, \Yobsj, \Yimpnotj, \theta)$
\end{enumerate}
\end{algorithmic}
\end{algorithm}

Under some regularity conditions the two algorithms are equivalent in finite samples if we can write $\pi(\theta) = \pi_j(\theta_j)\pi_{(-j)}(\theta_1,\theta_2,\dots,\theta_{j-1}, \theta_{j+1},\dots,\theta_p)$ for any $j$ and the set of $g_j$'s are compatible and correspond to the full conditionals of $f$ \citep{Hughes2014-au}.  This is sufficient to ensure that the conditional distributions in both steps of each algorithm agree.

If  $\pi(\theta) \neq \pi_j(\theta_j)\pi_{(-j)}(\theta_1,\theta_2,\dots,\theta_{j-1}, \theta_{j+1},\dots,\theta_p)$ for some $j$ but the conditional models are compatible  and correspond to the full conditionals of $f$, the two algorithms agree as $n\rightarrow\infty$ provided the FCS algorithm has a unique stationary distribution \citep{Liu2014-md}. Intuitively, in this case the data in $Y^{(-j)}$ influence $\theta_j$ indirectly through the other parameters, but the FCS algorithm ignores this information.  Asymptotically the priors become irrelevant in regular parametric models, but in finite samples inference based on the FCS imputations may be inefficient in this regime \citep{Seaman2016-mz}. 

Finally, \cite{Liu2014-md} show that if the FCS algorithm uses an inconsistent set of models but has a unique stationary distribution then MI estimates computed using imputations from Algorithm~\ref{alg:fcs} are consistent provided that the following conditions hold:
\begin{enumerate}
\item The collection of conditional models are incompatible, but become compatible with a joint model $f$ after constraining $\theta$.
\item The model class defined by $f$ contains the true distribution that generated the data.
\end{enumerate}
These are rather restrictive; verifying a unique stationary distribution is challenging, as is checking condition 1 above. It also seems unlikely that condition 2 will hold exactly for the simple parametric models in common use. 
\cite{Zhu2015-fn} provide some further convergence results for FCS algorithms where each observation is missing at most one value, but without assuming a unique stationary distribution for the FCS chain.

\subsection{Joint specifications: Sequential approach}

Sequential approaches to imputation modeling fix a permutation of $1,2,\dots,p$ and build up a joint distribution from a series of univariate models. For example, if the variables are already in the desired order we would have
\begin{equation}
f(Y) = f_1(Y_1)f_2(Y_2\mid Y_1)f_3(Y_3\mid Y_2, Y_1)\dots f_p(Y_p\mid Y_{p-1},\dots,Y_1).
\end{equation}
Examples of this approach include \citep{ibrahimlips, ibrahim, ibrahim:chen:lip:herr,Lee2016-wc,Xu2016-xy}, among others. 

Provided that each $f_j$ is a proper univariate probability model, a sequential specification always defines a coherent joint model, unlike FCS approaches. However, different orderings will generally lead to different joint distributions and potentially different fits. Heuristics have been proposed for selecting the order, for example ordering variables by their types (e.g. \cite{ibrahim}) or percentage of missing values (e.g., \cite{rubin1990efficiently}). 
The latter is particularly well-motivated when the missing data are monotone (when there is an ordering such that $R_{ij}=0\Rightarrow R_{ij'}=0$ for $j'>j$. ). If the missing data are not exactly monotone one can identify a permutation that is nearly monotone and use FCS or delete observed values to ``monotonize'' the missing data pattern, so that proper sequential techniques can be used for the majority of missing values (as in \cite{Rubin2003-uf} and extended in \cite{Li2014-lx}).

\begin{figure}

{\includegraphics[width=.8\linewidth]{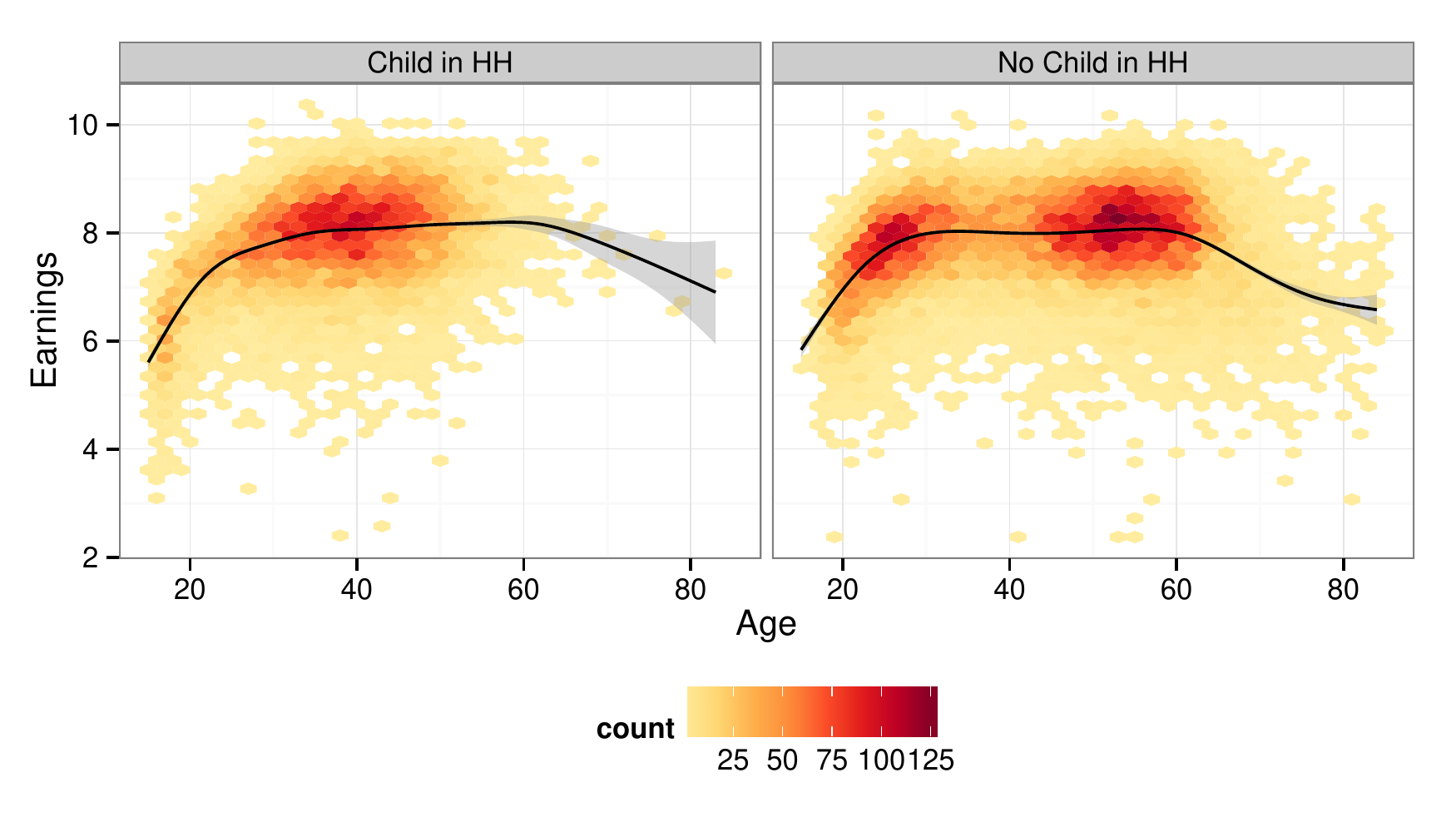}
}
\caption{Joint distribution of householder age and log total earnings, stratified on whether the household includes one of the householder's own children, using the population  \cite{Murray2016-cy} constructed from complete cases in the first wave of the Survey of Income and Program Participation's 2008 panel.}
\label{fig:sipp}
\end{figure}

Another consideration in joint-sequential modeling is that variables 
early in the sequence may have complex distributions because they are marginalized over many related covariates. For example, Figure~\ref{fig:sipp} shows the joint distribution of householder earnings and age, conditional on whether the householder has any children living in the same household (the data are from complete cases in wave one of the Survey of Income and Program Participation's 2008 panel). The distributions are quite complicated, and it would be difficult to capture them well with simple parametric regression models in any order.

\section{Choosing and assessing an imputation strategy}\label{sec:choosing}

\subsection{Comparing FCS and Joint approaches}
FCS and joint approaches have competing strengths. FCS models are relatively simple to implement and widely available in software, especially compared to joint-sequential approaches. Joint-simultaneous models including the multivariate normal, log-linear models, and the GLOM are also easy to set up and widely available, but inflexible in practice even relative to simple FCS procedures (e.g. \cite{VanBuuren2007, Stuart2009,He2010,Drechsler2010,kropko2014multiple}). 

More sophisticated joint models can be challenging to implement, although this is changing -- many of the nonparametric Bayesian methods have publicly available implementations (Appendix~\ref{sec:software}). However, even with a good implementation the nonparametric Bayesian models are generally more computationally expensive than simpler joint models (especially those based on low-rank methods, e.g. \cite{Audigier2016-va,Audigier2017-ye}) or FCS methods.  Joint-sequential approaches currently take more effort to set up, but they inherit many of the positive features of FCS and joint-simultaneous approaches (univariate models that are readily assessed and modified but also consistent with joint models). 

The convergence properties of FCS in general settings is still mostly an open question. The behavior of FCS algorithms under non- or quasi-Bayesian imputation procedures like PMM is entirely an open question.  While the lack of a coherent joint distribution does undermine the theoretical justifications for MI inference detailed in \cite{Rubin1987}, experience with FCS in simulations and real applications does not seem to suggest that either lack of convergence or compatibility with a joint model are necessarily overriding concerns.  

In fact, under the current theoretical results ensuring that the imputations generated by FCS converge to the imputations under a proper joint model requires using restrictive (implicit) joint models and there is strong empirical evidence that these joint models can be too simple to perform well with realistic data (e.g. \cite{Murray2016-cy,Akande2017-gq}).  Therefore at this point it would probably be a mistake to choose the models in an FCS imputation routine to try to ensure convergence; it seems much more important to use flexible, adaptive imputation models wherever possible, whether using a joint or FCS imputation strategy.

Imputers who do choose to use FCS should use flexible univariate models wherever possible and take care to assess apparent convergence of the algorithm, for example by computing traces of pooled estimates or other statistics and using standard MCMC diagnostics \citep[Chpater 11]{gelman2013bayesian}. It may also be helpful to examine the results of many independent runs of the algorithm with different initializations and to use random scans over the $p$ variables to try to identify any convergence issues and mitigate possible order dependence.

\subsection{Practical considerations derived from MI theory}
We can also compare methods on the practical considerations derived from theoretical results as summarized in Section~\ref{sec:implications}:

\subsubsection{Accounting for uncertainty.} Most of the methods reviewed above include some mechanism for reflecting imputation model uncertainty. Bayesian or approximately Bayesian methods (including the approximate Bayesian bootstrap) do this naturally, whether part of a joint modeling or FCS imputation routine.  Their behavior is not well understood in the FCS setting, however. Tree-based methods seem promising for some applications, but more work is required to find parameter settings and resampling strategies that make them reliably proper.

 \subsubsection{ Include as many variables as possible.} Joint-sequential models may be easier to fit than FCS with many covariates, since all but one univariate model will include fewer than $p$ predictors. Simultaneous joint models somewhat lag behind sequential and FCS approaches here. This is particularly true with mixed data types and many fully observed covariates -  most of these models are not easily adapted to condition on additional covariates, so fully observed variables must be included as additional variables in the joint model. Modeling fully observed variables instead of conditioning on them can waste ``degrees of freedom'' and lead to poorer model fit for the conditional distribution of the missing data. Carefully constructed models can help \citep{DeYoreo2016}, but seem to only go so far. 

 \subsubsection{ Use flexible imputation models.} Non- and semiparametric methods (Bayesian and otherwise, such as sequential tree-based methods) are flexible in their ability to capture certain unanticipated features of the data. Empirically these methods can outperform existing default MI procedures in simulations, particularly when the simulations are not built around simple parametric models themselves.  More of these realistic evaluations are needed, as discussed in Sections~\ref{sec:empiricalcomp} and ~\ref{sec:conclusion}.

However, with flexible imputation models it can be challenging to manually adjust the imputation model to incorporate prior information or address model misfit. Incorporating meaningful prior information into nonparametric Bayesian imputation models is challenging but not impossible; see e.g. \cite{Schifeling2016-zs} for a strategy to include prior information in DP-MPMN models. While iterative imputation model refinement and assessment is ideal, it is not always possible. Empirical evidence suggests that flexible imputation models are much better as defaults than simple parametric models or PMM using linear models.

\subsection{Empirical comparisons between methods}\label{sec:empiricalcomp}

Empirical comparisons of several different imputation models on realistic datasets are relatively rare. Most papers introducing a new imputation model evaluate it using synthetic data generated from a researcher-specified multivariate probability model. The new imputation model is typically compared to a small number of competitors.  These simulation studies can be informative -- for example, both \cite{Burgette2010-dq} and \cite{Doove2014-na} found evidence that imputations for continuous values generated via recursive partitioning can preserve interactions but underestimate main effects. However, models that are easy to simulate from and present in a paper will naturally be gross simplifications of the distribution of data in real populations. %

Simulations based on repeated sampling from realistic populations can be more informative.  In these studies a  population is compiled from existing data. Random samples are taken from these populations and values are ``blanked out'' via a known stochastic nonresponse mechanism. Each of the resulting incomplete datasets are multiply imputed and used to compute a range of estimates and confidence intervals, assessing the bias, coverage and efficiency of the MI estimates under the imputation model. Since the missing values are known, these can all be compared against the frequentist operating characteristics of the complete data procedure without appeal to asymptotic theory or other approximations.  While the results are specific to a particular population and a set of estimands, this framework is much closer to reality than fully synthetic examples.

There are several recent examples of this kind of evaluation: \cite{Akande2017-gq} compared FCS with CART, the DP-MPMN model described in \ref{sec:npbayesimp}, 
and a default application of FCS with main effects multinomial logistic regression in a large repeated-sampling study of imputation using categorical data from the American Community Survey. The DP-MPMN imputations tended to yield better coverage than FCS-CART overall, but had much worse coverage for a small number of estimands. \cite{Manrique-VallierSMTrunc} also demonstrated the utility of accounting for structural zeros in this model with a population constructed from publicly available data from the U.S. Census.  A default version of \cite{Murray2016-cy}'s joint model for mixed data types outperformed FCS using the default settings in R's \texttt{mice} package %
\citep{VanBuuren2011} in a large repeated-sampling study with data from the Survey of Income and Program Participation. Evidence suggested that misspecification bias was primarily to blame for FCS's poor performance. %

\subsection{Imputation model diagnostics}

A more obvious way to choose between imputation models is by fitting multiple and choosing the one that appears to fit the data best. Checking the fit of imputation models is challenging, but some approaches have been proposed. For methods that employ univariate regressions, imputers can examine standard diagnostics for those models \citep{Abayomi2008-if,su2011multiple}. \cite{Abayomi2008-if} suggested other diagnostic plots comparing imputed and observed values, primarily comparing marginal and bivariate distributions. Under MAR the distribution of missing values may be different than the distribution of observed values; \cite{Bondarenko2016-dj} used estimated response propensities to adjust for this and make diagnostic plots more comparable. \cite{He2012-ac} proposed posterior predictive checks, comparing the distribution of estimands computed on the multiply imputed datasets to the distribution of those estimands computed on entirely synthetic datasets generated by the imputation method (see also \cite{Nguyen2015-jy}).  These checks require the imputer to choose relevant estimands and generate many samples from posterior predictive distributions, which can be computationally expensive. 

\section{Conclusion}\label{sec:conclusion}

Over thirty years after Rubin's extensive treatment of MI \citep{Rubin1987}, experience with the method has cemented its reputation as a principled and practical solution to missing data problems. MI remains an active and fertile research area. While the behavior of the MI estimates have been the subject of intense scrutiny, relatively little is known about the comparative merits of various imputation models that have been proposed in recent years. Considerations based on theoretical findings suggest the use of more flexible imputation models where possible. Empirical evidence also suggests that simple defaults (MVN/log-linear models, or default FCS imputation using simple imputation models such as PMM with linear mean functions or regression models including only main effects) should be avoided, or at least carefully scrutinized.  

Nonparametric Bayesian methods for generating imputations have recently emerged as a promising technique for generating imputations. In addition to new model development, more work is needed on scalable posterior computation with these models. In addition, the heuristic justification for why Bayesian MI ``tends to be proper'' is based on the asymptotic behavior of parametric Bayesian models \citep{Rubin1987}. It would be interesting to revisit this argument from the perspective of Bayesian nonparametric models, where the asymptotics are more involved (see \cite{Rousseau2016-dw} for a recent review). For example, can semiparametric Bernstein von-Mises results be derived for likely targets of MI inference under Bayesian nonparametric models used for imputation?

Joint-sequential approaches appear understudied and underutilized in the literature, perhaps because they currently require more intervention to set up.  More research is needed on the implications of choosing different permutations of the variables in joint-sequential approaches. Further development of algorithmic approaches for selecting good joint-sequential variable orderings in the same vein as \cite{Li2014-lx} would also be welcome.  There remains considerable work to be done in characterizing the behavior of FCS approaches to generating imputations; while some theoretical results exist, they are limited in scope and do not address some of the most effective variants of these algorithms (including PMM and CART).

More empirical comparisons of imputation methods and models are also needed. The field  would benefit greatly from a repository of ready-to-use synthetic populations constructed from real data files.  A common set of samples from these populations complete with missing values already generated would allow for easy comparisons across methods. A forward-thinking statistical agency could kickstart this repository, providing a public good (and possibly improving the state of their own missing data imputation routines) by sponsoring an imputation challenge in the spirit of a Kaggle competition.  

The applications of MI have grown far beyond imputing item missing data in public use files: MI is used with synthetic data for disclosure limitation \citep{Rubin1993-pj,Reiter2002-lt, Raghunathan2003-yi}, to adjust for measurement error \citep{Cole2006-uf, Blackwell2015-eu}, and to perform statistical matching/data fusion \citep{Rassler2004-pb, Reiter2012-oa,Fosdick2016-fi}. In these new settings the amount of missing data can be much greater than typical applications of MI for item missing data, and imputation model development, selection, and assessment is even more consequential. We expect that new models and methods for multiple imputation will be an active research area for the foreseeable future.

\bibliographystyle{./imsart-nameyear}
\bibliography{paperpile,2013-08-mixedimp}

\appendix

\section{Software for Multiple Imputation}\label{sec:software}

Pointers to many software implementations of MI methods are available at \url{http://www.stefvanbuuren.nl/mi/Software.html}, an updated version of Appendix A of \cite{van2012flexible}.  As of December 2017, it is missing links to R packages for several nonparametric Bayesian joint models: These include the R packages MixedDataImpute (imputation for mixed continuous and categorical missing values using the model in \cite{Murray2016-cy}), NPBayesImpute (imputation for multivariate categorical data, possibly with structural zeros, as presented in \cite{Si2013,Manrique-Vallier2014-ct,Manrique-VallierSMTrunc}), and NestedCategBayesImpute (imputation got multivariate categorical data with hierarchical data structures, as described in \cite{Hu2017-mc}).

\end{document}